\documentclass[%
 reprint,
superscriptaddress,
 amsmath,amssymb,
 aps,
 prl
]{revtex4-1}

\usepackage{graphicx}% Include figure files
\usepackage{dcolumn}% Align table columns on decimal point
\usepackage{bm}% bold math
\usepackage{color}
\usepackage{hyperref}% add hypertext capabilities
\usepackage[mathlines]{lineno}% Enable numbering of text and display math
\newcommand{\Lc}{$\Lambda_c$}
\newcommand{\Lcpm}{$\Lambda_c^{\pm}$}
\newcommand{\sNN}{$\sqrt{s_{_{\rm NN}}}$}
\begin{document}

\preprint{APS/123-QED}

\title{First measurement of $\Lambda_c$ baryon production in Au+Au collisions at $\sqrt{s_{_{\rm NN}}}$ = 200 GeV}% Force line breaks with \\

%\collaboration{STAR Collaboration}%\noaffiliation
\affiliation{Abilene Christian University, Abilene, Texas   79699}
\affiliation{AGH University of Science and Technology, FPACS, Cracow 30-059, Poland}
\affiliation{Alikhanov Institute for Theoretical and Experimental Physics NRC "Kurchatov Institute", Moscow 117218, Russia}
\affiliation{Argonne National Laboratory, Argonne, Illinois 60439}
\affiliation{American University of Cairo, New Cairo 11835, New Cairo, Egypt}
\affiliation{Brookhaven National Laboratory, Upton, New York 11973}
\affiliation{University of California, Berkeley, California 94720}
\affiliation{University of California, Davis, California 95616}
\affiliation{University of California, Los Angeles, California 90095}
\affiliation{University of California, Riverside, California 92521}
\affiliation{Central China Normal University, Wuhan, Hubei 430079 }
\affiliation{University of Illinois at Chicago, Chicago, Illinois 60607}
\affiliation{Creighton University, Omaha, Nebraska 68178}
\affiliation{Czech Technical University in Prague, FNSPE, Prague 115 19, Czech Republic}
\affiliation{Technische Universit\"at Darmstadt, Darmstadt 64289, Germany}
\affiliation{ELTE E\"otv\"os Lor\'and University, Budapest, Hungary H-1117}
\affiliation{Frankfurt Institute for Advanced Studies FIAS, Frankfurt 60438, Germany}
\affiliation{Fudan University, Shanghai, 200433 }
\affiliation{University of Heidelberg, Heidelberg 69120, Germany }
\affiliation{University of Houston, Houston, Texas 77204}
\affiliation{Huzhou University, Huzhou, Zhejiang  313000}
\affiliation{Indian Institute of Science Education and Research (IISER), Berhampur 760010 , India}
\affiliation{Indian Institute of Science Education and Research (IISER) Tirupati, Tirupati 517507, India}
\affiliation{Indian Institute Technology, Patna, Bihar 801106, India}
\affiliation{Indiana University, Bloomington, Indiana 47408}
\affiliation{Institute of Modern Physics, Chinese Academy of Sciences, Lanzhou, Gansu 730000 }
\affiliation{University of Jammu, Jammu 180001, India}
\affiliation{Joint Institute for Nuclear Research, Dubna 141 980, Russia}
\affiliation{Kent State University, Kent, Ohio 44242}
\affiliation{University of Kentucky, Lexington, Kentucky 40506-0055}
\affiliation{Lawrence Berkeley National Laboratory, Berkeley, California 94720}
\affiliation{Lehigh University, Bethlehem, Pennsylvania 18015}
\affiliation{Max-Planck-Institut f\"ur Physik, Munich 80805, Germany}
\affiliation{Michigan State University, East Lansing, Michigan 48824}
\affiliation{National Research Nuclear University MEPhI, Moscow 115409, Russia}
\affiliation{National Institute of Science Education and Research, HBNI, Jatni 752050, India}
\affiliation{National Cheng Kung University, Tainan 70101 }
\affiliation{Nuclear Physics Institute of the CAS, Rez 250 68, Czech Republic}
\affiliation{Ohio State University, Columbus, Ohio 43210}
\affiliation{Institute of Nuclear Physics PAN, Cracow 31-342, Poland}
\affiliation{Panjab University, Chandigarh 160014, India}
\affiliation{Pennsylvania State University, University Park, Pennsylvania 16802}
\affiliation{NRC "Kurchatov Institute", Institute of High Energy Physics, Protvino 142281, Russia}
\affiliation{Purdue University, West Lafayette, Indiana 47907}
\affiliation{Rice University, Houston, Texas 77251}
\affiliation{Rutgers University, Piscataway, New Jersey 08854}
\affiliation{Universidade de S\~ao Paulo, S\~ao Paulo, Brazil 05314-970}
\affiliation{University of Science and Technology of China, Hefei, Anhui 230026}
\affiliation{Shandong University, Qingdao, Shandong 266237}
\affiliation{Shanghai Institute of Applied Physics, Chinese Academy of Sciences, Shanghai 201800}
\affiliation{Southern Connecticut State University, New Haven, Connecticut 06515}
\affiliation{State University of New York, Stony Brook, New York 11794}
\affiliation{Temple University, Philadelphia, Pennsylvania 19122}
\affiliation{Texas A\&M University, College Station, Texas 77843}
\affiliation{University of Texas, Austin, Texas 78712}
\affiliation{Tsinghua University, Beijing 100084}
\affiliation{University of Tsukuba, Tsukuba, Ibaraki 305-8571, Japan}
\affiliation{United States Naval Academy, Annapolis, Maryland 21402}
\affiliation{Valparaiso University, Valparaiso, Indiana 46383}
\affiliation{Variable Energy Cyclotron Centre, Kolkata 700064, India}
\affiliation{Warsaw University of Technology, Warsaw 00-661, Poland}
\affiliation{Wayne State University, Detroit, Michigan 48201}
\affiliation{Yale University, New Haven, Connecticut 06520}

\author{J.~Adam}\affiliation{Brookhaven National Laboratory, Upton, New York 11973}
\author{L.~Adamczyk}\affiliation{AGH University of Science and Technology, FPACS, Cracow 30-059, Poland}
\author{J.~R.~Adams}\affiliation{Ohio State University, Columbus, Ohio 43210}
\author{J.~K.~Adkins}\affiliation{University of Kentucky, Lexington, Kentucky 40506-0055}
\author{G.~Agakishiev}\affiliation{Joint Institute for Nuclear Research, Dubna 141 980, Russia}
\author{M.~M.~Aggarwal}\affiliation{Panjab University, Chandigarh 160014, India}
\author{Z.~Ahammed}\affiliation{Variable Energy Cyclotron Centre, Kolkata 700064, India}
\author{I.~Alekseev}\affiliation{Alikhanov Institute for Theoretical and Experimental Physics NRC "Kurchatov Institute", Moscow 117218, Russia}\affiliation{National Research Nuclear University MEPhI, Moscow 115409, Russia}
\author{D.~M.~Anderson}\affiliation{Texas A\&M University, College Station, Texas 77843}
\author{A.~Aparin}\affiliation{Joint Institute for Nuclear Research, Dubna 141 980, Russia}
% ...

\collaboration{STAR Collaboration}\noaffiliation

\date{\today}% It is always \today, today,
             %  but any date may be explicitly specified

\begin{abstract}
We report on the first measurement of the charmed baryon \Lcpm~production at midrapidity ($|y|$\,$<$\,1) in Au+Au collisions at \sNN~= 200\,GeV collected by the STAR experiment at the Relativistic Heavy Ion Collider. The \Lc/$D^0$ (denoting ($\Lambda_c^++\Lambda_c^-$)/($D^0+\overline{D}^0$)) yield ratio is measured to be 1.08 $\pm$ 0.16 (stat.)\, $\pm$ 0.26 (sys.) in the 0--20\% most central Au+Au collisions for the transverse momentum ($p_T$) range 3 $<$ $p_T$ $<$ 6\,GeV/$c$. This is significantly larger than the PYTHIA model calculations for $p+p$ collisions. The measured \Lc/$D^0$ ratio, as a function of $p_T$ and collision centrality, is comparable to the baryon-to-meson ratios for light and strange hadrons in Au+Au collisions. Model calculations including coalescence hadronization for charmed baryon and meson formation reproduce the features of our measured \Lc/$D^0$ ratio.
\end{abstract}

% PACS, the Physics and Astronomy
                             % Classification Scheme.
%\keywords{Suggested keywords}%Use showkeys class option if keyword
                              %display desired
\maketitle

%\tableofcontents

Heavy ion collisions offer a unique opportunity to study Quantum Chromodynamics (QCD), the theory describing strong interactions between quarks and gluons through color charges. Data collected from the Relativistic Heavy Ion Collider (RHIC) and the Large Hadron Collider (LHC) demonstrate that a novel QCD matter, Quark-Gluon Plasma (QGP), in which quarks and gluons are deconfined, is created in high-energy nucleus-nucleus collisions~\cite{Akiba:2015jwa,Adams:2005dq,*Adcox:2004mh,*Back:2004je,*Arsene:2004fa}. QCD hadronization is a nonperturbative process and remains a challenging process to model. Fragmentation fractions measured in high energy $ee$, $ep$ and $pp$ collisions have been used to successfully describe hadron production at high transverse momentum ($p_T$), and are deployed in Monte Carlo (MC) event generators like PYTHIA~\cite{Sjostrand:2006za} using a string fragmentation hadronization scheme. Recently, different schemes, such as color reconnection (CR) in PYTHIA, where strings from different multi-parton interactions are allowed to recombine, have been developed to reproduce the low-$p_T$ hadron data, including an enhanced production of baryons, in $pp$ collisions~\cite{Bierlich:2015rha}. In central heavy-ion collisions, baryon-to-meson ratios for light and strange hadrons in 2\,$<$\,$p_T$\,$<$\,6\,GeV/$c$ show an enhancement compared to $pp$ collisions~\cite{Abelev:2006jr,Abelev:2013xaa,Abelev:2007ra}. A coalescence hadronization mechanism, in which hadrons can be formed via recombination of close-by partons in phase space in the deconfined QGP, has been utilized to describe the enhancement in heavy-ion collisions~\cite{Lin:2003jy,Fries:2008hs}. Alternatively to these microscopic schemes, a statistical hadronization scheme, which determines hadron yields statistically by their quantum numbers and thermal properties of the system, is used to fit successfully various light and strange hadron integrated yields in $ee$, $pp$ and heavy-ion collisions~\cite{Wheaton:2004qb}.

Due to their large masses, heavy quarks ($c$, $b$) are predominately created from initial hard scatterings in heavy-ion collisions. The relative yields of heavy-flavor hadrons can serve as a tag to study their hadronization process. The $c$ quark fragmentation fraction ratio ($c$$\rightarrow$$\Lambda_c^+$)/($c$$\rightarrow$$D^0$) was measured to be around 0.10--0.15 in $ee$ and $ep$ collisions~\cite{Barate:1999bg,Abramowicz:2013eja,Lisovyi:2016qjn}. Recently, ALICE and LHCb measured~\cite{Acharya:2017kfy,Aaij:2018iyy} the $\Lambda_c/D^0$ ratio in $p+p$ and $p+\mathrm{Pb}$ collisions at the LHC to be 0.4--0.5 at 2\,$<$\,$p_T$\,$<$\,8\,GeV/$c$, larger than the PYTHIA model calculation based on string fragmentation. PYTHIA model with color reconnection yields a larger $\Lambda_c/D^0$ ratio that is close to the data~\cite{Acharya:2017kfy}.

In heavy-ion collisions, models including coalescence hadronization of charm quarks predict a large $\Lambda_c/D^0$ ratio of $\sim$\,1, in the low to intermediate $p_T$ regions ($<\sim$ 8 GeV/c)~\cite{Oh:2009zj,Greco:2003vf,Lee:2007wr}. The ALICE Collaboration reported the $\Lambda_c/D^0$ ratio to be $\sim$1 at 6\,$<$\,$p_T$\,$<$12\,GeV/$c$ in Pb+Pb collisions at $\sqrt{s_{_{\rm NN}}}$ = 5.02\,TeV, consistent with a contribution of coalescence hadronization for charm quarks~\cite{Acharya:2018ckj}. Measurement of \Lcpm~production in heavy-ion collisions over a broad momentum region, particularly at lower $p_T$, will offer significant insights into the hadronization mechanism of charm quarks in the presence of a QGP. Furthermore, understanding the hadronization mechanism of charm quarks in heavy-ion collisions is crucial to the study of charm quark energy loss in the QGP using the measurements of nuclear modification factors ($R_{\mathrm{AA}}$) of $D$ mesons~\cite{Adam:2018inb,Sirunyan:2017xss,Adam:2015sza} in heavy-ion collisions. Since the charm quarks are dominantly produced through initial hard scatterings, a large baryon-to-meson ratio directly impacts the charm meson $R_{\mathrm{AA}}$. 

In this Letter, we report on the first measurement of \Lcpm~production in Au+Au collisions at \sNN~= 200\,GeV. The analysis is carried out at midrapidity ($|y|$\,$<$\,1), and utilized a total of 2.3 billion minimum bias (MB) triggered events collected by the STAR experiment during 2014 and 2016 runs at RHIC. The Heavy Flavor Tracker (HFT)~\cite{Contin:2017mck}, a four-layer high resolution silicon detector, was used for excellent vertex resolution that improves significantly the signal-to-background ratio for charmed hadron reconstruction. The MB events are selected by requiring a coincidence between the east and west Vertex Position Detectors~\cite{Llope:2003ti}. The events are required to have the reconstructed primary vertex (PV) position along the beam direction within 6 cm from the detector center, to ensure good HFT acceptance. The collision centrality, a measure of the geometric overlap between the two colliding nuclei, is defined using the measured charged track multiplicity at mid-rapidity, as compared to a Monte Carlo Glauber simulation~\cite{Abelev:2008ab}.

The \Lcpm~baryons are reconstructed via the hadronic decay channel $\Lambda_c^{+}$ $\rightarrow K^-\pi^+ p$ and its charge conjugate. Charged particle tracks are reconstructed from hits in the STAR Time Projection Chamber (TPC)~\cite{Anderson:2003ur} and HFT detectors, in a 0.5 T magnetic field. Tracks are required to have a minimum of 20 TPC hits (out of a maximum of 45) and at least three hits in the HFT sub-detectors. The tracks are also required to be within pseudorapidity $|\eta|$\,$<$\,1 with $p_T$\,$>$\,0.5\,GeV/$c$. Particle identification (PID) is achieved by a combination of the ionization energy loss, $dE/dx$, measured by the TPC and the timing, measured by the Time Of Flight (TOF) detector~\cite{Llope:2012zz}.

\begin{figure}[htbp]
\center{
\includegraphics[width=0.9\columnwidth]{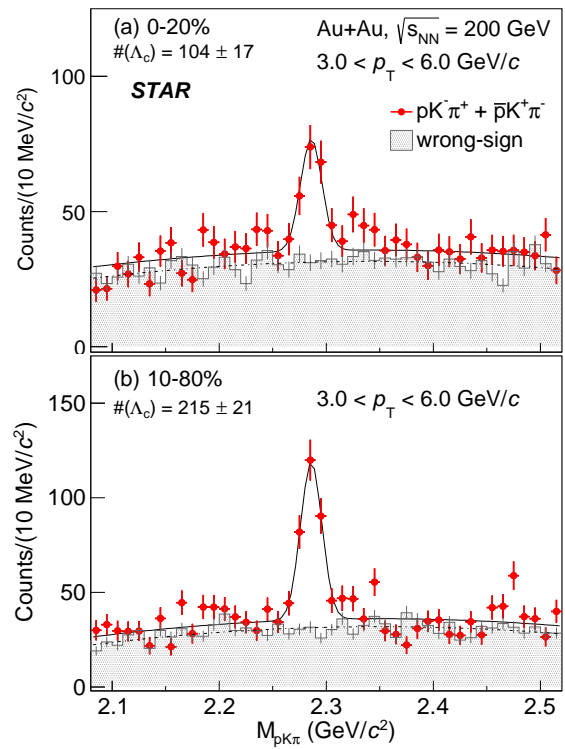}%
}
\caption{The $pK\pi$ invariant mass distributions for right-sign (solid red points) and wrong-sign (shaded histograms) combinations in Au+Au collisions at $\sqrt{s_{_{\rm NN}}}$ = 200\,GeV for 0--20\% (top) and 10--80\% (bottom) centrality classes. The wrong-sign distributions are scaled by 1/3, the ratio of number of right-sign to wrong-sign combinations for the $pK\pi$ triplet. The error bars shown are statistical uncertainties. The solid line depicts a fit with a Gaussian function, for $\Lambda_c^{\pm}$ signal, and a second order polynomial function, the shape of which is fixed by fit to the wrong-sign distribution (dashed line), for the background.
}
\label{fig:signal}
\end{figure}

The \Lcpm~decay vertex is reconstructed as the mid-point of the distance of closest approach (DCA) between the three daughter tracks. To improve separation of signal from combinatorial background of tracks originating from the primary vertex, we utilized a supervised machine learning algorithm, the Boosted Decision Trees (BDT), implemented in the TMVA package~\cite{Hocker:2007ht}. The BDTs are trained with a signal sample of \Lcpm $\rightarrow K\pi p$ decays simulated using the EvtGen generator~\cite{Lange:2001uf} with detector effects taken into account and a background sample of wrong-sign $K\pi p$ combinations from data. The variables characterizing the decay topology, viz. the decay length, DCA of daughter tracks to the PV and the DCA of the reconstructed $\Lambda_c$ candidate to PV are used as input variables in the training. The cut on BDT response is optimized for maximum \Lcpm~signal significance using the estimated number of signal and background \Lcpm~ candidates in data. Figure~\ref{fig:signal} shows examples of invariant mass distributions with the BDT selection, of $K\pi p$ triplets with the right and wrong-sign (scaled by 1/3) combinations. The distributions in the 0--20\% most central collisions (top) and the 10--80\% central collisions (bottom), the centrality range used for $p_{T}$-dependent measurement, are shown. The right-sign distributions are fit to a Gaussian for the signal plus a second order polynomial for the background, with the shape of the polynomial function fixed from fitting to the wrong-sign distribution. The raw signal yields are obtained as the counts of the right-sign triplets within a mass window of three standard deviations of the Gaussian fit with background counts, evaluated using the polynomial component of the fit in the same mass window, subtracted.

The \Lcpm~reconstruction efficiency is evaluated using a hybrid method, similarly to the $D^0$ spectra measurement with the STAR HFT~\cite{Adam:2018inb}. The TPC tracking efficiency is obtained using the standard embedding technique used in many other STAR analyses~\cite{Agakishiev:2011ar}. The PID efficiencies are evaluated using pure $\pi$, $K$, $p$ samples from data. The HFT tracking and the BDT selection efficiency are calculated using a data-driven simulation framework with the input distributions taken from the real data. The input distributions include the TPC-to-HFT matching efficiency (the fraction of good TPC tracks matched to hits in HFT) and the DCA distributions of tracks with respect to the reconstructed collision vertex.
Protons reconstructed in the real data have a sizable secondary contribution from other hyperon decays, which impacts the TPC-to-HFT matching ratio and DCA distributions.  
 A correction factor to the efficiency calculated using the data-driven simulation is evaluated using Au+Au events from HIJING~\cite{Gyulassy:1994ew} propagated through the STAR GEANT detector geometry~\cite{Agostinelli:2002hh} and digital signals embedded into those from zero-bias data (denoted HIJING+ZB). Zero-bias data consist of events taken with no collision requirement, and capture the background conditions in the detectors during the run. The $p_{\mathrm{T}}$ distributions of protons and hyperons from HIJING are reweighted to match data~\cite{Abelev:2006jr,Agakishiev:2011ar}. The events are then reconstructed with the same algorithm as the real data. The correction is calculated as a ratio of the efficiency from the data-driven simulation, using the input distributions for inclusive tracks from the reconstructed HIJING+ZB data, to the one using inputs from primary tracks from the same data. The correction factor is 
 found to be about 30\% with very weak $p_T$ and centrality dependences. The impact of the finite primary vertex resolution on the reconstruction efficiency obtained by this method is also evaluated using the HIJING+ZB events with procedures similar to those described in~\cite{Adam:2018inb}. It is found to be within 10\% for the 50--80\% centrality class and negligible for more central events.
The yields are finally corrected for the \Lcpm $\rightarrow K\pi p$ branching ratio (B.R.) of 6.28 $\pm$ 0.32\%~\cite{Tanabashi:2018oca}.

The systematic uncertainties to the measurement include the uncertainties in raw yield extraction and various efficiency correction factors. The former is evaluated by varying the background estimation method (varying the fit range, choice of background function and leaving the background shape unconstrained), and is between 6--14\% in the measured $p_T$ region. The contribution to the yield under the mass peak from incorrectly assigned PID for daughter tracks is less than 1\%. The TPC efficiency uncertainty is evaluated to be $\sim$15\%, and PID efficiency uncertainties to be $\sim$6\%, for three daughter tracks combined. The uncertainty in the HFT tracking and topological cut efficiency is estimated by changing the BDT response cuts so that the reconstruction efficiency varies by 50\% above and below relative to the nominal one. The resulting non-statistical variations to final results are included in the systematic uncertainties and range from 10--15\%. For the correction factor due to secondary protons, the uncertainties from the measured proton and $\Lambda$ spectra~\cite{Abelev:2006jr,Agakishiev:2011ar}, as well as those on other hadrons that decay to protons, are propagated. This uncertainty is estimated to be about 4\%. We also include a 10\% uncertainty from a closure test for the data-driven simulation method, evaluated by comparing the efficiencies calculated using data-driven simulation with input distributions from reconstructed HIJING+ZB events, to the efficiencies evaluated directly from the reconstructed HIJING+ZB events.
The feed-down contribution from bottom hadrons to the measurements is found to be small and less than 4\% in the measured $p_{T}$ range. Finally, the uncertainty in the decay B.R. from the latest PDG~\cite{Tanabashi:2018oca} value is added as a global normalization uncertainty in the \Lcpm~yield.

\begin{table}[h!]
\centering
\small{\begin{tabular}{c|c}
\hline
\hline
$p_{\mathrm{T}}$ (GeV/$c$)      & 1/(2$\pi p_{\mathrm{T}}N_{\mathrm{evt}}$)$\mathrm{d}^{\mathrm{2}}N$/d$p_{\mathrm{T}}$d$y$ (GeV/$c$)$^{-2}$\\[2pt] %\tabularnewline
\hline
2.5 - 3.5 & 8.2$\times \mathrm{10}^{\mathrm{-4}} \pm$ 1.4$\times \mathrm{10}^{\mathrm{-4}}$ (stat.) $\pm$ 2.4$\times \mathrm{10}^{\mathrm{-4}}$ (sys.)\\[2pt]
3.5 - 5.0 & 6.0$\times \mathrm{10}^{\mathrm{-5}} \pm$ 7.7$\times \mathrm{10}^{\mathrm{-6}}$ (stat.) $\pm$ 1.5$\times \mathrm{10}^{\mathrm{-5}}$ (sys.)\\[2pt]
5.0 - 8.0 & 2.1$\times \mathrm{10}^{\mathrm{-6}} \pm$ 3.8$\times \mathrm{10}^{\mathrm{-7}}$ (stat.) $\pm$ 5.5$\times \mathrm{10}^{\mathrm{-7}}$ (sys.)\\[2pt]
\hline
\hline
\end{tabular}}\normalsize
\caption{\label{fig:fig3} The \Lcpm~invariant yields measured in the 10-80\% centrality class for the different $p_{\mathrm{T}}$ bins, in Au+Au collisions at \sNN = 200 GeV.}
\label{tab:spectra}
\vspace{-8pt}
\end{table}

\begin{figure}[htbp]
\center{
\includegraphics[width=0.9\columnwidth]{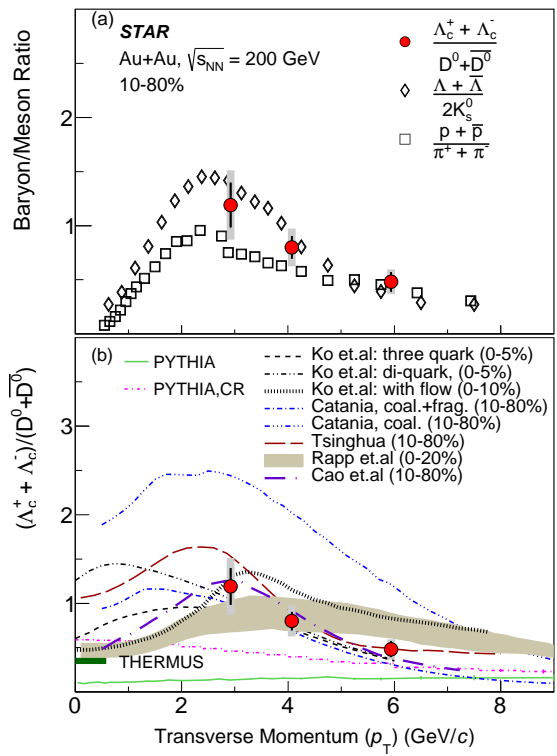}%
}
\caption{The measured \Lc/$D^0$ ratio at midrapidity ($|y|<$ 1) as a function of $p_{\rm T}$ for Au+Au collisions at \sNN~= 200\,GeV in 10-80\% centrality, compared to the baryon-to-meson ratios for light and strange hadrons (top) and various model calculations (bottom). The vertical lines and shaded boxes on the \Lc/$D^0$ data points indicate statistical and systematic uncertainties respectively. The $p_{\rm T}$ integrated \Lc/$D^0$ ratio from the THERMUS~\cite{Wheaton:2004qb} model calculation with a freeze-out temperature of $T_{\rm ch}=160$\,MeV is shown as a horizontal bar on the left axis of the plot.
}
\label{fig:ratioPt}
\end{figure}

The \Lcpm~invariant yields in the 10-80\% centrality class for the different $p_{\mathrm{T}}$ bins are shown in Table~\ref{tab:spectra}, along with the statistical and systematic uncertainties. The 10-80\% centrality class is chosen for $p_{\mathrm{T}}$-dependent measurement as it had the best $\Lambda_c$ signal significance in the measured regions. The ratio of the invariant yield of \Lcpm~to that of $D^{0}$ is shown as a function of $p_T$ in Fig.~\ref{fig:ratioPt} for the 10--80\% centrality class. The correlated systematic uncertainties from efficiency correction that go into both \Lcpm~ and $D^0$ measurements, cancel. Figure~\ref{fig:ratioPt} (a) compares the $\Lambda_c/D^0$ ratio to the baryon-to-meson ratios from light and strange-flavor hadrons~\cite{Agakishiev:2011ar,Abelev:2006jr}. The $\Lambda_c/D^0$ ratio is comparable in magnitude to the $\Lambda/K^0_s$ and $p$/$\pi$ ratios and shows a similar $p_T$ dependence in the measured region.

The measured values are compared to different model calculations in panel (b) of Fig.~\ref{fig:ratioPt}. The values show a significant enhancement compared to the calculations from the latest PYTHIA 8.24 release (Monash tune~\cite{Skands:2014pea}) without CR~\cite{Bierlich:2015rha}. The implementation with CR (mode2 in ~\cite{Bierlich:2015rha}) enhances the baryon production with respect to mesons and gives a $\Lambda_c/D^0$ yield ratio consistent with those measured in $p$+$p$ and $p$+$\mathrm{Pb}$ collisions at the LHC~\cite{Acharya:2017kfy,Lhcb:2018ppb}. However, both calculations fail to fully describe the Au+Au data and its $p_{T}$ dependence. The mode without CR is ruled out at a $p$-value of 1 $\times \mathrm{10}^{\mathrm{-4}}$ ($\chi^2$/NDF = 20.7/3), while the CR mode gives a $p$-value of 0.04 ($\chi^2$/NDF = 8.2/3) using a reduced $\chi^2$ test. %The measured $\Lambda_c/D^0$ yield ratio in Au+Au collisions at $\sqrt{s_{_{\rm NN}}}$ = 200\,GeV is also larger compared to those measured in $p$+$p$ and $p$+$\mathrm{Pb}$ collisions at the LHC~\cite{Acharya:2017kfy,Lhcb:2018ppb}. 

Figure~\ref{fig:ratioPt} (b) shows the comparison to calculations from various models that include coalescence hadronization of charm quarks (labelled Ko et.al: with three quarks and diquarks~\cite{Oh:2009zj}, Ko et.al: with flow~\cite{Cho:2019lxb}, Catania~\cite{Plumari:2017ntm}, Tsinghua~\cite{Zhao:2018jlw}, Rapp et.al~\cite{He:2019vgs} and Cao et.al~\cite{Cao:2019iqs}). The models differ among themselves in the choice of hadron wave functions, light and charm quark spectra in the QGP and also treatment of space-time correlations during coalescence and excited states that decay into $\Lambda_c$ and $D^0$ that are considered. Most of the models are able to give enhanced $\Lambda_c/D^0$ yield ratios and describe the measured $p_{\mathrm{T}}$ dependence of the ratio. A reduced $\chi^2$ test is carried out, taking into account the finite $p_{\mathrm{T}}$ bin-width in the measurement. The Catania model calculations of the $\Lambda_c/D^0$ ratio from hadrons formed only through coalescence hadronization over-predict the measurement at all $p_{\mathrm{T}}$ (reduced $\chi^2$ = 26.1). The calculations from Ko et al. with flow give a reduced $\chi^2$ value of 4.8, mainly from the over-prediction of the ratio in the highest two $p_{\mathrm{T}}$ bins. The other coalescence model calculations are consistent with data within uncertainties over the measured $p_{\mathrm{T}}$ range. It should be noted that the calculations from Rapp et al. and Ko et al. have different centrality ranges than in the measurement, which may impact the $\chi^2$ values quoted. In the models discussed above, charm quark radial flow is implicitly included mainly through the charm quark diffusion in the medium. However, it was found that a purely radial flow effect without coalescence hadronization, evaluated using a Blast-Wave model with freeze-out parameters from $D^0$ measurement~\cite{Adam:2018inb}, causes the $\Lambda_c/D^0$ ratio to rise strongly with increasing $p_{T}$ in the measured $p_{T}$ region. This is similar to the behavior observed for light hadrons~\cite{Abelev:2013xaa}, and opposite to the trend measured in the data. The comparisons favor coalescence hadronization as having an important role in charm-quark hadronization in the presence of QGP. The data offer constraints to the model parameters and to the coalescence probabilities of charm quarks in the medium. 

%The models differ among themselves in the choice of hadron wave functions, light and charm quark spectra in the QGP and also treatment of space-time correlations during coalescence and excited states that decay into $\Lambda_c$ and $D^0$ that are considered. 

The $p_T$-integrated $\Lambda_c/D^{0}$ ratio is calculated to be 0.80 $\pm$ 0.12 (stat)\, $\pm$ 0.22 (sys,data) $\pm$ 0.41 (sys,model). The coalescence model curves shown in Fig.~\ref{fig:ratioPt}(b) were used to extrapolate to $p_{T}$\,=\,0 GeV/c, with the mean of the extrapolated values from different models taken as the central value and the maximum difference between them included in the systematic uncertainty. The ratio is consistent, including extrapolation uncertainties, to the value (0.35) from thermal model calculation using THERMUS~\cite{Wheaton:2004qb} with a freeze-out temperature, $T_{\rm ch}$ = 160\,MeV. This suggests \Lcpm~contribute sizably to the total charm yield in heavy-ion collisions.

\begin{figure}[htbp]
\center{
\includegraphics[width=0.9\columnwidth]{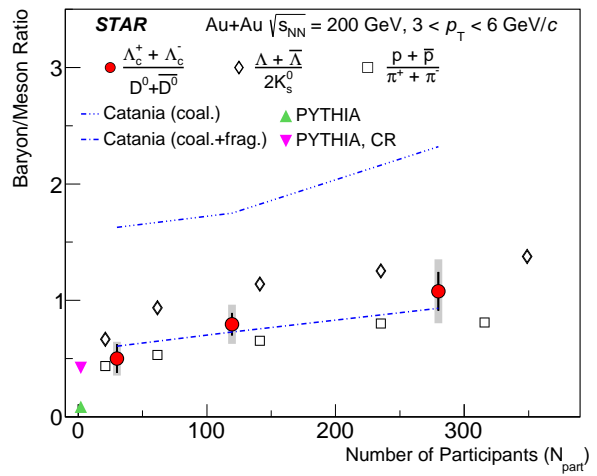}%
}
\caption{The measured \Lc/$D^0$ yield ratio in 3 $< p_{\mathrm{T}} <$ 6\,GeV/$c$ (solid circles) as a function of collision centrality (expressed in $N_{\rm part}$) for Au+Au collisions at \sNN~= 200\,GeV. The open diamonds and squares show the baryon-to-meson ratio measured for strange and light-flavor hadrons respectively. The vertical lines and the shaded boxes on the \Lc/$D^0$ data points indicate statistical and systematic uncertainties respectively. The dashed curves indicate the \Lc/$D^0$ ratio calculated from a model with charm quark coalescence, and the up and down triangles indicate the ratios from the PYTHIA model for $p+p$ collisions without and with color reconnection (CR) respectively, for the same $p_{\rm T}$ region.
}
\label{fig:ratioNpart}
\end{figure}

The centrality dependence of the $\Lambda_c/D^0$ ratio, plotted as function of the number of participant nucleons $N_{\mathrm{part}}$, for 3\,$<$\,$p_T$\,$<$\,6\,GeV/$c$ is shown in Fig.~\ref{fig:ratioNpart}. The measurements correspond to the centrality ranges 50-80\%, 20-50\% and 0-20\%. The $\Lambda_c/D^0$ ratio shows an increase towards more central collisions. The increasing trend is qualitatively similar to that seen for the baryon-to-meson ratio for light and strange-flavor hadrons, and to that predicted by coalescence model calculations. The measured $\Lambda_c/D^0$ ratio in 0-20\% central collisions of 1.08 $\pm$ 0.16(stat.) $\pm$ 0.26(sys.) is larger than the values from PYTHIA 8.2 without CR (at 3.1 $\sigma$ significance) and with CR (at 2.1 $\sigma$ significance). 

In summary, STAR reports on the first measurement of \Lcpm~baryon production in Au+Au collisions at \sNN~= 200\,GeV utilizing its high-resolution silicon detector. The measured \Lc/$D^0$ yield ratio at midrapidity ($|y|$\,$<$\,1) is found to be comparable to the baryon-to-meson ratios for light and strange-flavor hadrons in the same kinematic regions. The large \Lc/$D^0$ ratio also suggests that charmed baryons contribute significantly to the total charm cross section at midrapidity in heavy-ion collisions at RHIC. The \Lc/$D^0$ ratio in Au+Au collisions is considerably larger than the PYTHIA expectation at the same energy. Several model calculations that include coalescence hadronization for charm hadron formation can reproduce the features of our data.
Our data is expected to offer significant constraints towards the understanding of QCD hadronization in the finite temperature region, and to the charm quark transport and energy loss in the QGP.

We thank the RHIC Operations Group and RCF at BNL, the NERSC Center at LBNL, and the Open Science Grid consortium for providing resources and support.  This work was supported in part by the Office of Nuclear Physics within the U.S. DOE Office of Science, the U.S. National Science Foundation, the Ministry of Education and Science of the Russian Federation, National Natural Science Foundation of China, Chinese Academy of Science, the Ministry of Science and Technology of China and the Chinese Ministry of Education, the National Research Foundation of Korea, Czech Science Foundation and Ministry of Education, Youth and Sports of the Czech Republic, Hungarian National Research, Development and Innovation Office, New National Excellency Programme of the Hungarian Ministry of Human Capacities, Department of Atomic Energy and Department of Science and Technology of the Government of India, the National Science Centre of Poland, the Ministry  of Science, Education and Sports of the Republic of Croatia, RosAtom of Russia and German Bundesministerium fur Bildung, Wissenschaft, Forschung and Technologie (BMBF) and the Helmholtz Association.

\bibliography{Lambda_c}

\end{document}